\documentclass[epj]{svjour}
\usepackage{graphicx}
\begin{document}
\title{Biased diffusion in a piecewise linear random potential}
\author{S.I. Denisov\inst{1,2,}\thanks{\rm{e-mail: stdenis@pks.mpg.de}},
E.S. Denisova\inst{2} \and H. Kantz\inst{1}}
\institute{Max-Planck-Institut f\"{u}r Physik komplexer Systeme, N\"{o}thnitzer
Stra{\ss}e 38, D-01187 Dresden, Germany \and Sumy State University, 2
Rimsky-Korsakov Street, 40007 Sumy, Ukraine }
\date{Received:  }
\abstract{We study the biased diffusion of particles moving in one direction
under the action of a constant force in the presence of a piecewise linear
random potential. Using the overdamped equation of motion, we represent the
first and second moments of the particle position as inverse Laplace
transforms. By applying to these transforms the ordinary and the modified
Tauberian theorem, we determine the short- and long-time behavior of the
mean-square displacement of particles. Our results show that while at short
times the biased diffusion is always ballistic, at long times it can be either
normal or anomalous. We formulate the conditions for normal and anomalous
behavior and derive the laws of biased diffusion in both these cases.
}

\maketitle

\section{Introduction}
\label{Intr}

The biased diffusion, i.e., diffusion accompanied by the directional transport
of diffusing objects (particles), exists in many systems subjected to external
force fields. In systems with quenched disorder it can be naturally described
by the Langevin equation in which disorder is accounted by time-independent
random potentials and thermal fluctuations are modeled by white noise. Because
of its relative simplicity and efficiency, the Langevin-based approach provides
an important tool for studying the transport properties in disordered media
\cite{BG}. In particular, wit\-hin this framework a number of effects arising
from the joint action of quenched disorder and thermal fluctuations, including
some features of biased diffusion, has been successfully studied for particles
moving under a constant force in a one-dimensional random potential
\cite{Sch,DV,PKDK,GB,DH,LV,Mon,RE}.

Since in the most cases considered earlier the distribution of the random force
which corresponds to a given random potential has unbounded support, particles
cannot be transported to an arbitrary large distance if thermal fluctuations
are absent. Put differently, in the noiseless case particles remain localized
in a finite region at all times. It is clear that delocalization can occur only
if the above mentioned distribution has bounded support. At this condition, the
biased diffusion can be caused by a time-periodic external force or a constant
one. In both these cases the diffusive behavior of particles results solely
from quenched disorder, but the directional transport has different nature.
Indeed, a time-periodic force induces the directional transport due to the
ratchet effect which exists in random potentials of a special class, i.e.,
ratchet potentials with quenched disorder \cite{PASF,GLZH,ZLAF,DLDHK}. In
contrast, a constant force (if it exceeds a critical value) induces the
directional transport in arbitrary potentials due to the direct action on
particles. In this case particles move only in one direction and, as a
consequence, the completely anisotropic case of biased diffusion, when the
probability of motion along and against the external force equals 1 and 0,
respectively, is realized. Some features of this diffusion exhibiting normal
behavior were considered in \cite{KLS,DKDH1,DKDH2} within a general approach
based on the equation of motion of diffusing particles. At the same time,
anomalous regimes of this diffusion were studied only in the framework of
continuous time random walk \cite{DK}, which adequately describes the long-time
case.

In this paper we develop a \textit{unified} approach for the study of biased
diffusion of particles moving under a constant force in a piecewise linear
random potential. It is based on the overdamped equation of motion and gives a
possibility to get the diffusion laws for both short and long times. The paper
is organized as follows. In Section \ref{sec:Prob} we describe the model and
derive the probability density of the particle position in terms of the
probability density of the residence time. In Section \ref{sec:Mom} we
represent the first two moments of the particle distribution as inverse Laplace
transforms. The biased diffusion at short and long times is studied in Sections
\ref{sec:Short} and \ref{sec:Long}, respectively, using the ordinary and the
modified Tauberian theorem for the Laplace transform. Specifically, we obtain
the law of diffusion at short times (Sections \ref{sec:Short}), formulate the
conditions for normal and anomalous diffusion at long times (Section
\ref{sec:Cond}), calculate the diffusion coefficient and analyze its dependence
on the driving force (Section \ref{sec:Norm}), and derive the laws of anomalous
diffusion (Section \ref{sec:Anom}). Finally, in Section \ref{sec:Concl} we
summarize our results.

\section{Probability density of the particle position}
\label{sec:Prob}

\begin{figure}
    \centering
    \includegraphics[totalheight=4cm]{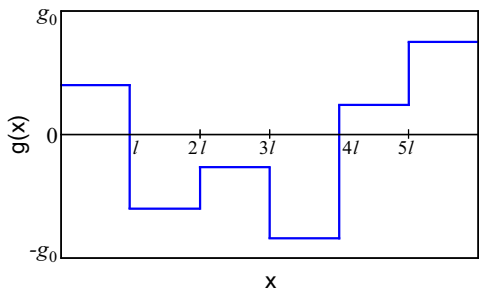}
    \caption{\label{fig.1} Sample path of the piecewise constant
    random force $g(x)$ that corresponds to a given realization of
    the piecewise linear random potential (\ref{U(x)}).}
\end{figure}

We consider the unidirectional motion of a particle which occurs under the
action of a constant driving force $f (>0)$ in a piecewise linear random
potential $U(x)$. At $x\geq 0$ we define this potential as follows:
\begin{equation}
    U(x) = -(x-nl)g^{(n)} +U(nl), \qquad x \in [nl, nl+l),
    \label{U(x)}
\end{equation}
where $n=0,1,\ldots$ and the slopes $g^{(n)}$ are assumed to be independent
random variables having the same probability density function $u(g)$. Next we
suppose that this density function is symmetric, i.e., $u(-g) = u(g)$, and has
bounded support, i.e., $g \in [-g_{0}, g_{0}]$. If the condition $U(nl) =
-l\sum_{m=0}^{n-1} g^{(m)}$ holds for $n\geq 1$ ($U(0)$ is an arbitrary
constant which can be chosen to be zero) then the potential $U(x)$ is
continuous and the corresponding piecewise constant random force, $g(x)=-dU
(x)/dx$, takes the form $g(x) = g^{(n)}$ if $x$ belongs to the $n$th interval
$[nl, nl+l)$ (see figure \ref{fig.1}). Thus, neglecting the inertial effects,
we can describe the particle dynamics by the overdamped equation of motion
\begin{equation}
    \nu \dot{X}_{t} = f + g(X_{t}),
    \label{eq motion}
\end{equation}
where $X_{t}$ ($X_{0}=0$) is the particle position and $\nu$ is the damping
coefficient.

According to the above assumptions, at $f\in (0,g_{0})$ particles cannot be
transported to an arbitrary large distance in the positive direction of the
axis $x$. On the contrary, they are stopped at some distance $L=l N$ from the
origin, where $N$ is the number of that interval which is characterized by the
conditions $f>-g^{(n)}$ (for all $n<N$) and $f\leq -g^{(N)}$. This number
depends on the sample paths of $g(x)$ and its mean $\langle N \rangle$ (the
angular brackets denote an average over these sample paths) can be easily
calculated by introducing the probability $I=\int_{-g_{0}}^{f}dg\,u(g)$ that
$g(x)<f$. Indeed, since the random forces $g^{(m)}$ are statistically
independent on different intervals, the probability $W_{n}$ that $N=n$ can be
written as $W_{n} = I^{n} \int_{f}^{g_{0}}dg\,u(g)$, i.e., $W_{n} = I^{n} -
I^{n+1}$. With this expression for $W_{n}$ we obtain $\langle N \rangle =
\sum_{n=1}^ {\infty} nW_{n} = \sum_{n=1}^{\infty} I^{n}$, and the use of the
geometric series formula $\sum_{n=0}^{\infty}I^{n} = 1/(1-I)$ yields $\langle N
\rangle = I/(1-I)$. As a consequence, the average distance $\langle L \rangle$
to the point where particles are stopped is given by
\begin{equation}
    \langle L \rangle = l\,\frac{I}{1-I}.
    \label{aver}
\end{equation}
Thus, if $f<g_{0}$, i.e., $I<1$, then the final state of particles is their
localization at a finite average distance (\ref{aver}). We note also that if
the probability density $u(g)$ has no $\delta$ singularities at $g= \pm g_{0}$,
i.e., $u(g)$ does not contain the terms proportional to $\delta (g-g_{0})$ and
$\delta (g+g_{0})$, where $\delta(\cdot)$ is the Dirac $\delta$ function, then
$I\to 1$ and $\langle L \rangle \to \infty$ as $f\to g_{0}$.

In contrast, if $f>g_{0}$ then $f+g(X_{t})>0$ and, according to the motion
equation (\ref{eq motion}), $X_{t} \to \infty$ as $t\to \infty$. As it follows
from (\ref{aver}), the same behavior holds for $f=g_{0}$ as well. In other
words, at $f\geq g_{0}$ particles can be transported to an arbitrary large
distance along the positive direction of the $x$-axis. In this cases the
distribution of particles on the positive semi-axis $x$ is described by the
probability density
\begin{equation}
    P(x,t) = \langle \delta(x-X_{t}) \rangle
    \label{defP}
\end{equation}
that $X_{t} =x$. Assuming that $X_{t} = nl + X_{t}^{(n)}$ and $X_{t}^{(n)} \in
[0,l)$, it is convenient to rewrite the probability density of the particle
position in the form
\begin{equation}
    P(x,t) = \sum_{n=0}^{\infty}P_{n}(x,t)
    \label{Pa}
\end{equation}
with $P_{n}(x,t) = \langle \delta(x - nl - X_{t}^{(n)}) \rangle$. In order to
calculate $P_{n}(x,t)$, let us first introduce the residence time $\tau^{(n)}$
of a particle in the $n$th interval $[nl, nl+ l)$, i.e., time that a particle
spends moving from the point $nl$ to the point $nl+l$. Since in the overdamped
regime the motion occurs with a constant velocity $v^{(n)} =l/\tau^{(n)}$, we
obtain $X_{t}^{(n)} = l(t - \sum_{m=0}^{n-1}\tau^{(m)})/\tau^{(n)}$ if $t -
\sum_{m=0}^{n-1}\tau^{(m)} \in [0,\tau^{(n)})$ and
\begin{equation}
    P_{n}(x,t) = \left\langle \delta\! \left [x - nl - \frac{l}
    {\tau^{(n)}}\! \left(t - \sum_{m=0}^{n-1}\tau^{(m)}\right)
    \right] \right\rangle.
    \label{Pn}
\end{equation}
According to (\ref{eq motion}), the residence time is given by $\tau^{(n)} =
\nu l/(f + g^{(n)})$. Because of the properties of $g^{(n)}$, these times in
different intervals are statistically independent and distributed with the same
probability density $p(\tau)$. Taking into account that $\tau\in [\tau_{
\rm{min}}, \tau_{\rm{max}}]$ with
\begin{equation}
    \tau_{\rm{min}} = \frac{\nu l}{f+g_{0}}, \qquad
    \tau_{\rm{max}} = \frac{\nu l}{f-g_{0}},
    \label{tau}
\end{equation}
from the condition $p(\tau) |d\tau| = u(g) |dg|$ we can express the probability
density $p(\tau)$ of the residence time through the probability density $u(g)$
of the random force:
\begin{equation}
    p(\tau) = \left\{ \begin{array}{ll}
    \displaystyle \frac{\nu l}{\tau^{2}}\,
    u\!\left(\frac{\nu l}{\tau} -f\right),
    & \tau\in [\tau_{\rm{min}},\tau_{\rm{max}}]
    \\ [10pt]
    0,
    & \rm{otherwise}.
    \end{array}
    \right.
    \label{p}
\end{equation}

In turn, the quantities $P_{n}(x,t)$ can be expressed thro\-ugh $p(\tau)$.
Specifically, from the definition (\ref{Pn}) at $n=0$ we obtain
\begin{equation}
    P_{0}(x,t) = \int_{t}^{\infty}d\tau_{0} p(\tau_{0})\delta\!
    \left(x- \frac{l}{\tau_{0}}\,t\right).
    \label{P_0}
\end{equation}
It is also not difficult to see that $P_{n}(x,t)$ at $n\geq 1$ is given by
\begin{eqnarray}
    P_{n}(x,t) &=& \int_{0}^{t}d\tau_{0}p(\tau_{0})
    \int_{0}^{t-\tau_{0}}d\tau_{1}p(\tau_{1}) \ldots
    \int_{0}^{t-\sum_{m=0}^{n-2}\tau_{m}}
    \nonumber\\[6pt]
    && \times d\tau_{n-1} p(\tau_{n-1}) \int_{t-
    \sum_{m=0}^{n-1}\tau_{m}}^{\infty}d\tau_{n} p(\tau_{n})
    \nonumber\\[6pt]
    && \times \delta\!\left[x-nl - \!\frac{l}{\tau_{n}}\!\left(t-
    \sum_{m=0}^{n-1}\!\tau_{m}\right)\!\right].
    \label{P_n2}
\end{eqnarray}
The last two formulas can be written in a more compact form using the notation
$v(t)*w(t)$ for the convolution of two functions $v(t)$ and $w(t)$,
\begin{equation}
    v(t)*w(t) = \int_{0}^{t}d\tau\, v(\tau) w(t-\tau).
    \label{con}
\end{equation}
With this definition, we can introduce the $n$-fold convolution of the
probability density of the residence time as follows: $p^{*n}(t)=
p^{*n-1}(t)*p(t)$. Assuming also that $p^{*0}(t) = \delta(t)$, we obtain
$P_{0}(x,t) = p^{*0}(t)* P_{0}(x,t)$, $P_{n}(x,t) = p^{*n}(t)*P_{0} (x-nl,t)$,
and so the probability density of the particle position can be represented in
the form
\begin{equation}
    P(x,t)= \sum_{n=0}^{\infty}p^{*n}(t)*P_{0}(x-nl,t).
    \label{P}
\end{equation}
We note that there is a possibility of further simplification of $P(x,t)$ by
using the Laplace transform method. However, the representation (\ref{P}) is
quite sufficient for our purpose, i.e., finding the moments of $X_{t}$.

\section{Moments of the particle position}
\label{sec:Mom}

The $k$th moment of the particle position is defined in the usual way:
\begin{equation}
    \langle X^{k}_{t} \rangle = \int_{0}^{\infty} dx x^{k}P(x,t)
    \qquad (k=1,2,\ldots).
    \label{kth}
\end{equation}
Substituting (\ref{P}) into (\ref{kth}) and using (\ref{P_0}), for the first
moment we obtain
\begin{equation}
    \langle X_{t} \rangle = l \sum_{n=0}^{\infty}p^{*n}(t)*
    \int_{t}^{\infty}d\tau(n+t/\tau)p(\tau).
    \label{1a}
\end{equation}
Because of the convolution structure of the summands in (\ref{1a}), it is
reasonable to apply the Laplace transform defined as
\begin{equation}
    h_{s} = \mathcal{L}\{ h(t) \} = \int_{0}^{\infty}dt e^{-st}h(t)
    \label{Lap}
\end{equation}
($\rm{Re}\,s>0$) to both sides of this formula. Using the convolution theorem
for the Laplace transform, $\mathcal{L}\{ v(t)*w(t) \} = v_{s}w_{s}$, this
yields
\begin{equation}
    \langle X_{t} \rangle_{s} = l \sum_{n=0}^{\infty}p^{n}_{s}\,
    \mathcal{L}\left\{\int_{t}^{\infty}d\tau(n+t/\tau)p(\tau)\right\}.
    \label{1sa}
\end{equation}

By changing the order of integration in the Laplace transform of the integral
term, we obtain
\begin{eqnarray}
    &\displaystyle\mathcal{L}\left\{\int_{t}^{\infty}d\tau(n+t/\tau)
    p(\tau)\right\}  = \int_{0}^{\infty}d\tau p(\tau) \quad\;\;&
    \nonumber\\[6pt]
    &\displaystyle \times \int_{0}^{\tau} dt(n+t/\tau) e^{-st}
    = n\, \frac{1-p_{s}}{s} - \frac{p_{s}}{s} + F_{s},&
    \label{rel1a}
\end{eqnarray}
where
\begin{equation}
    F_{s} = \frac{1}{s^{2}}\int_{0}^{\infty}d\tau\, \frac{1-e^{-s\tau}}
    {\tau}\,p(\tau).
    \label{F}
\end{equation}
Therefore, taking into account that $\sum_{n=0}^{\infty} p^{n}_{s}=1/
(1-p_{s})$ and $\sum_{n=1}^{\infty} np^{n}_{s} = p_{s}/(1- p_{s})^{2}$, formula
(\ref{1sa}) reduces to
\begin{equation}
    \langle X_{t} \rangle_{s} = l\, \frac{F_{s}}{1-p_{s}}.
    \label{1sb}
\end{equation}
Finally, applying to (\ref{1sb}) the inverse Laplace transform defined as
\begin{equation}
    h(t) = \mathcal{L}^{-1}\{ h_{s} \} = \frac{1}{2\pi i} \int_{c- i\infty}^
    {c+i\infty}ds\, e^{st}h_{s}
    \label{InvLap}
\end{equation}
(the real number $c$ is chosen to be larger than the real parts of all
singularities of $h_{s}$), we find the first moment in the form
\begin{equation}
    \langle X_{t} \rangle = l\,\mathcal{L}^{-1}\!\left\{ \frac{F_{s}}
    {1-p_{s}} \right\}.
    \label{1b}
\end{equation}

The second moment,
\begin{equation}
    \langle X_{t}^{2} \rangle = l^{2} \sum_{n=0}^{\infty}p^{*n}(t)*
    \int_{t}^{\infty}d\tau(n+t/\tau)^{2}p(\tau),
    \label{2a}
\end{equation}
can also be determined by the method of Laplace transform. The similar
calculations lead to
\begin{equation}
    \langle X_{t}^{2} \rangle_{s} = l^{2} \sum_{n=0}^{\infty}
    p^{n}_{s}\, \mathcal{L}\left\{\int_{t}^{\infty}d\tau
    (n+t/\tau)^{2}p(\tau) \right\},
    \label{2sa}
\end{equation}
where
\begin{eqnarray}
    \mathcal{L}\left\{\int_{t}^{\infty}d\tau(n+t/\tau)^{2}p(\tau)
    \right\} &=& n^{2} \frac{1-p_{s}}{s} + 2n\frac{sF_{s} - p_{s}}{s}
    \nonumber\\[6pt]
    &&- \frac{p_{s}}{s} + G_{s}
    \label{rel2}
\end{eqnarray}
with
\begin{equation}
    G_{s} =\frac{2}{s^{3}}\int_{0}^{\infty}d\tau\,\frac{1-(1+s\tau)
    e^{-s\tau}}{\tau^{2}}\,p(\tau).
    \label{G}
\end{equation}
Carrying out summation over $n$ in equation (\ref{2sa}) [for this purpose we
use the above expressions for the series and the formula $\sum_{n=1}^{\infty}
n^{2}p^{n}_{s} = p_{s}(1+ p_{s})/(1- p_{s})^{3}$], we obtain
\begin{equation}
    \langle X_{t}^{2} \rangle_{s} = l^{2} \frac{2p_{s}F_{s} +
    (1-p_{s})G_{s}} {(1-p_{s})^{2}}
    \label{2sb}
\end{equation}
and so
\begin{equation}
    \langle X_{t}^{2} \rangle = l^{2}\mathcal{L}^{-1}\!\left\{
    \frac{2p_{s}F_{s} +(1-p_{s})G_{s}} {(1-p_{s})^{2}} \right\}.
    \label{2b}
\end{equation}
As it will be shown below, the representations  (\ref{1b}) and (\ref{2b}) of
the moments $\langle X_{t} \rangle$ and $\langle X^{2}_{t} \rangle$ are very
useful for finding the short- and long-time behavior of the mean-square
displacement of a particle, or the variance of the particle position, defined
as
\begin{equation}
    \sigma^{2}_{X}(t) = \langle X^{2}_{t} \rangle - \langle X_{t}
    \rangle^{2}.
    \label{def var}
\end{equation}

In conclusion of this section, we briefly discuss the differences between our
model and the continuous-time random walk (CTRW) models. Introduced many years
ago \cite{MW}, the CTRW approach has become one of the most important tools for
the study of anomalous transport (see e.g. \cite{BG}, \cite{Hug,AH,MK,Z,KRS}).
In the simplest version of the CTRW is assumed that the random waiting time,
i.e., time that a walking particle waits before it jumps to another position,
and the random jump size are statistically independent. Since equation (\ref{eq
motion}) describes the continuous motion of a particle, this so-called jump
model is in general inapplicable to our case. Due to physical reasons, much
attention has also been paid to the case in which the waiting time (considered
as the motion time) and the jump size (considered as the passed distance) are
coupled in such a way that a walking particle moves with a constant velocity
\cite{SWK,MLW,ZK,Bar,SM}. However, because of the last condition, this velocity
model also does not provide an adequate description of the biased diffusion in
a piecewise linear random potential. From this point of view, the CTRW model
\cite{ZSS} in which a walking particle moves with a random velocity could be
the most appropriate. But in \cite{ZSS} is assumed that the random velocity and
motion time are statistically independent, while in our case the particle
velocity $v^{(n)} = (f+g^{(n)})/\nu$ and the residence time $\tau^{(n)} = \nu
l/ (f+g^{(n)})$ are perfectly dependent, i.e., $v^{(n)} \tau^{(n)} = l$.

Thus, none of the CTRW models completely describes the considered type of
biased diffusion. Nevertheless, if the waiting time is associated with the
residence time and the jump size with $l$ then the long-time behavior of
$X_{t}$ can be adequately described by the CTRW jump model \cite{DK}. It is
therefore reasonable to compare the behavior of $X_{t}$ with that following
from the unidirectional CTRW on a semi-infinite chain of period $l$. Within
this approach, the particle position is described by a discrete variable $Y_{t}
= lN(t)$, where $N(t)$ is the random number of jumps up to time $t$, and the
waiting time is characterized by the same probability density $p(\tau)$. Using
the moments of $N(t)$ that are known from the theory of CTRWs (see e.g.
\cite{Hug}), the first two moments of $Y_{t}$ can be written as
\begin{equation}
    \langle Y_{t} \rangle = l\mathcal{L}^{-1}\!\left\{ \frac{p_{s}}
    {s(1-p_{s})} \right\}, \;\; \langle Y^{2}_{t} \rangle = l^{2}
    \mathcal{L}^{-1}\!\left\{\frac{p_{s}^{2} + p_{s}}{s(1-p_{s})^{2}}
    \right\}\!.
    \label{1,2}
\end{equation}
The difference in the time dependence of the variances $\sigma^{2}_{X}(t)$ and
$\sigma^{2}_ {Y}(t) =\langle Y^{2}_{t} \rangle -\langle Y_{t} \rangle^{2}$ will
be clarified in the next sections.

\section{Short-time behavior of the variance}
\label{sec:Short}

The calculation of the inverse Laplace transforms in formulas (\ref{1b}) and
(\ref{2b}) is a difficult technical problem. Fortunately, in the short- and
long-time limits this problem can be avoided. Such a possibility provides the
celebrated Tauberian theorem for the Laplace transform \cite{Hug,Fel} which
states that if $h(t)$ is a monotonic function and
\begin{equation}
    h_{s} = \mathcal{L}\{ h(t) \} \sim L\!\left( \frac{1}{s} \right)\!
    \frac{1}{s^{\rho}}
    \label{hs1}
\end{equation}
as $s \to 0$ ($s\to \infty$) then
\begin{equation}
    h(t) = \mathcal{L}^{-1}\{ h_{s} \} \sim \frac{1}{\Gamma(\rho)}\,
    L (t)\,t^{\rho-1}
    \label{ht1}
\end{equation}
as $t \to \infty$ ($t\to 0$). Here, $\rho \in (0,\infty)$, $\Gamma(x)$ is the
gamma function, and $L(t)$ is a slowly varying function at infinity (zero),
i.e., $L(\lambda t) \sim L(t)$ as $t\to \infty$ ($t\to 0$) for all $\lambda>0$.
It should be stressed that, in contrast to the Laplace transform (\ref{Lap}),
the parameter $s$ in (\ref{hs1}) is assumed to be a positive real number. We
note also that if $s\to 0$ then the Tauberian theorem is valid in a more
general case, namely, if $h(t)$ is ultimately monotone, i.e., monotone on the
interval $(t',\infty)$ for some $t'>0$.

Thus, for finding the short-time behavior of the moments $\langle X_{t}
\rangle$ and $\langle X_{t}^{2} \rangle$ at $f\geq g_{0}$ we should determine
the leading terms of the Laplace transforms $\langle X_{t} \rangle_{s}$ and
$\langle X_{t}^{2} \rangle_{s}$ as $s$ tends to infinity. Taking into account
that $\tau \in [\tau_{\rm{min}}, \tau_{\rm{max}}]$ and $p_{s} \to 0$ as $s\to
\infty$, from (\ref{1sb}) and (\ref{2sb}) in this limit we obtain
\begin{equation}
    \langle X_{t} \rangle_{s} \sim \frac{l}{s^{2}} \int_{\tau_
    {\rm{min}}}^{\tau_{\rm{max}}} d\tau\, \frac{p(\tau)}{\tau},
    \quad
    \langle X_{t}^{2} \rangle_{s} \sim \frac{2l^{2}}{s^{3}}\int_
    {\tau_{\rm{min}}}^{\tau_{\rm{max}}}d\tau\, \frac{p(\tau)}{\tau^{2}}.
    \label{X1,2s}
\end{equation}
Using (\ref{p}) and the symmetry property of the probability density of the
random force, $u(-g) = u(g)$, the integrals in (\ref{X1,2s}) can be expressed
as
\begin{equation}
    \int_{\tau_{\rm{min}}}^{\tau_{\rm{max}}}d\tau\, \frac{p(\tau)}
    {\tau} = \frac{f}{\nu l},
    \quad
    \int_{\tau_{\rm{min}}}^{\tau_{\rm{max}}}d\tau\, \frac{p(\tau)}
    {\tau^{2}} = \frac{f^{2}+\sigma^{2}_{g}}{(\nu l)^{2}},
    \label{int1,2}
\end{equation}
where $\sigma^{2}_{g} = \int_{-g_{0}}^{g_{0}}dg g^{2}u(g)$ is the variance of
the random force $g(x)$. Therefore, as follows from the Tauberian theorem, the
short-time behavior of the first two moments is described by the asymptotic
formulas
\begin{equation}
    \langle X_{t} \rangle \sim \frac{f}{\nu}\,t,
    \quad
    \langle X_{t}^{2} \rangle \sim \frac{f^{2} + \sigma^{2}_{g}}
    {\nu^{2}}\,t^{2}.
    \label{X1,2}
\end{equation}
They show that at small times ($t\ll \nu l/f$) and $f\geq g_{0}$ the biased
diffusion is always ballistic, i.e., $\sigma^{2}_{X}(t)$ is proportional to
$t^{2}$:
\begin{equation}
    \sigma^{2}_{X}(t) \sim \frac{\sigma^{2}_{g}}{\nu^{2}}\,t^{2}.
    \label{Var1}
\end{equation}
We note that the diffusion law (\ref{Var1}) also follows from the motion
equation (\ref{eq motion}). Indeed, the first two moments of the particle
position $X_{t} \sim t(f + g^{(0)})/\nu$, which is the solution of this
equation at $t \to 0$, are the same as in (\ref{X1,2}).

For finding the short-time behavior of the moments of the discrete variable
$Y_{t}$ the Tauberian theorem cannot be used. The reason is that $\langle Y_{t}
^{k} \rangle_{s}$ at $s \to \infty$ tends to zero more rapidly than any
positive power of $1/s$ (since $p_{s} \propto e^{-s\tau_{\rm{min}}}$ as $s \to
\infty$). Nevertheless, exact conclusions about the short-time behavior of
$\langle Y_{t} ^{k} \rangle$ can be drown directly from the definition $\langle
Y_{t} ^{k} \rangle = l^{k} \sum_{n=1}^{\infty}n^{k} \mathcal{P} (n,t)$, where
$\mathcal{P}(n,t)$ is the probability that $N(t)=n$. Indeed, accounting for the
fact that $\mathcal{P}(n,t)= p(t)*\mathcal{P} (n-1,t)$ ($n \geq 1$) and
$\mathcal{P}(0,t) =\int_{t}^{\infty} d\tau p(\tau)$ \cite{Hug}, we make sure
(see also (\ref{p}) and (\ref{con})) that if $t< \tau_{\rm{min}}$ then
$\mathcal{P}(0,t) = 1$ and $\mathcal{P}(n,t) = 0$ for all $n \geq 1$. Hence, at
$t< \tau_{ \rm{min}}$ all moments of $Y_{t}$ are equal to zero and so the
short-time behavior of $Y_{t}$  is qualitatively different from that for
$X_{t}$.

\section{Long-time behavior of the variance}
\label{sec:Long}

\subsection{Conditions of normal and anomalous diffusion}
\label{sec:Cond}

According to (\ref{p}), the $m$th moment of the residence time, $\overline{
\tau^{m}} = \int_{0}^{\infty} d\tau \tau^{m} p(\tau)$ $(m=1,2,\ldots)$, can be
written in the form
\begin{equation}
    \overline{\tau^{m}} = (\nu l)^{m} \int_{-g_{0}}^{g_{0}}dg\,
    \frac{u(g)}{(f+g)^{m}}.
    \label{mth1}
\end{equation}
Since the probability density $u(g)$ is properly normalized on the interval
$[-g_{0}, g_{0}]$, i.e., $\int_ {-g_{0}}^ {g_{0}} dg u(g) = 1$, the above
formula leads to the condition $\overline {\tau^{m}} \leq (\nu l)^{m}/(f
-g_{0})^{m}$. It shows that if $f>g_{0}$ then all moments $\overline{\tau^{m}}$
are finite. Since in this case $\overline{\tau^{2}}< \infty$, the classical
central limit theorem for sums of a random number of random variables \cite{GK}
is applied to $X_{t}$ and suggests that $\sigma^{2}_{X}(t) \propto t$ as $t \to
\infty$. In other words, at $f>g_{0}$ the biased diffusion of particles is
expected to be always normal.

In contrast, if $f= g_{0}$ then the residence time varies from $\tau_{\rm{min}}
= \nu l/2g_{0}$ to $\tau_{\rm {max}} = \infty$, and $\overline{\tau^{2}}$ can
be either finite or infinite depending on the asymptotic behavior of $p(\tau)$
as $\tau \to \infty$ (i.e., asymptotic behavior of $u(g)$ as $g \to -g_{0}$).
While in the former case the biased diffusion is also expected to be normal, in
the latter (when $\overline{\tau^{2}} = \infty$) the above mentioned central
limit theorem becomes inapplicable and one may expect that the biased diffusion
is anomalous. Next we assume that at $f=g_{0}$ the probability density
$p(\tau)$ of the residence time is characterized by the asymptotic formula
\begin{equation}
    p(\tau) \sim \frac{a}{\tau^{1+\alpha}}  \quad (\tau \to \infty)
    \label{asymp}
\end{equation}
with positive parameters $a$ and $\alpha$. According to it $\overline{\tau^{2}}
< \infty$ if $\alpha>2$ and $\overline{\tau^{2}} = \infty$ if $\alpha \in
(0,2]$. The parameters $a$ and $\alpha$ depend on the explicit form of the
probability density $u(g)$ of the piecewise constant random force $g(x)$ and
are in general not independent. In particular, for the symmetric beta
probability density
\begin{equation}
    u(g) = \frac{\Gamma(\beta + 1/2)}{g_{0}\sqrt{\pi}\,
    \Gamma(\beta)} \!\left( 1- \frac{g^{2}}{g_{0}^{2}}
    \right)^{\! \beta-1}
    \label{u(g)}
\end{equation}
($\beta>0$) we have $\alpha = \beta$ and
\begin{equation}
    a = \!\left( \frac{2\nu l}{g_{0}} \right)^{\! \alpha}
    \frac{\Gamma(\alpha + 1/2)}{2\sqrt{\pi}\,\Gamma(\alpha)}.
    \label{a}
\end{equation}

Thus, the biased diffusion is expected to be normal if $f>g_{0}$ or $f=g_{0}$
and $\alpha>2$, and anomalous if $f=g_{0}$ and $\alpha \in (0,2]$. In the next
two sections we validate these criteria  and derive the laws of normal and
anomalous diffusion.

\subsection{Normal biased diffusion}
\label{sec:Norm}

\subsubsection{Modified Tauberian theorem.}
\label{sec:Taub}

The ordinary Tauberian theorem formulated in (\ref{hs1}) and (\ref{ht1})
permits us to find only the leading terms of the asymptotic expansion of $h(t)$
at $t\to 0$ and $t\to \infty$. Therefore, in this form it can be used for
determining the long-time behavior of the variance $\sigma^{2}_{X}(t)$ only in
the case when the leading terms of $\langle X_{t} \rangle^{2}$ and $\langle
X^{2}_{t} \rangle$ are different (see section \ref{sec:Anom1}). But in order to
find $\sigma^{2}_{X}(t)$ in the case with $\langle X_{t} \rangle^{2} \sim
\langle X^{2}_{t} \rangle$, we need to know at least two leading terms of each
of the asymptotic expansions of $\langle X_{t} \rangle$ and $\langle X^{2}_{t}
\rangle$. With one exception (see section \ref{sec:Anom3}), these terms can be
found by applying the modified version of the Tauberian theorem. This version
follows directly from the ordinary one by replacing $h_{s}$ by $h_{s} -
r/s^{\eta}$ and using the exact result $\mathcal {L}^{-1} \{ 1/s^{\eta} \}=
t^{\eta-1}/ \Gamma (\eta)$ \cite{Erd}. According to this, we formulate the
modified Tauberian theorem as follows: If $h(t) - r t^{\eta-1}/ \Gamma (\eta)$
is ultimately monotone and
\begin{equation}
    h_{s} \sim \frac{r}{s^{\eta}} + L\!\left( \frac{1}{s} \right)\!
    \frac{1}{s^{\rho}}
    \label{hs2}
\end{equation}
($\eta > \rho >0$) as $s\to 0$ then
\begin{equation}
    h(t) \sim \frac{r}{\Gamma(\eta)}\,t^{\eta-1} + \frac{1}
    {\Gamma(\rho)}\, L (t)\,t^{\rho-1}
    \label{ht2}
\end{equation}
as $t\to \infty$.

It is important to emphasize that the terms $rt^{\eta-1}/\Gamma(\eta)$ and
$L(t) t^{\rho-1}/ \Gamma(\rho)$ represent the first two terms of the long-time
expansion of $h(t)$ only if $\eta>\rho$. The reason is that in the opposite
case (when $\eta \leq \rho$) the second term of the asymptotic expansion of
$\mathcal {L}^{-1} \{ L(1/s)/ s^{\rho} \}$ at $t\to \infty$ may not be
negligible in comparison with $\mathcal {L}^{-1} \{ r/ s^{\eta} \}$. In order
to illustrate this fact, we assume that $L(1/s) = \ln(1/s)$ and consider the
Laplace transform of the following form:
\begin{equation}
    h_{s} = \frac{1}{s^{\eta}} + \ln\!\left( \frac{1}{s} \right)\!
    \frac{1}{s^{\rho}}.
    \label{hs3}
\end{equation}
Since $\mathcal {L}^{-1}\! \{ \ln(1/s)/ s^{\rho} \} = t^{\rho-1}[\ln t -
\psi(\rho)]/\Gamma(\rho)$ \cite{Erd}, where $\psi(\rho)$ is the digamma
function, i.e., the logarithmic derivative of the gamma function
$\Gamma(\rho)$, the inverse Laplace transform of $h_{s}$ yields an exact result
\begin{equation}
    h(t) = \frac{t^{\eta-1}}{\Gamma(\eta)} + \frac{t^{\rho-1}}
    {\Gamma(\rho)}\, [\ln t - \psi(\rho)].
    \label{ht3}
\end{equation}
Keeping in the long-time limit two leading terms of $h(t)$, we obtain
\begin{equation}
    h(t) \sim \left\{ \begin{array}{ll}
    t^{\eta-1}/\Gamma(\eta) + t^{\rho-1} \ln t /\Gamma(\rho),
    & \eta > \rho \\ [6pt]
    t^{\eta-1}[\ln t - \psi(\eta) + 1]/\Gamma(\eta),
    & \eta = \rho \\ [6pt]
    t^{\rho-1}[\ln t - \psi(\rho)]/\Gamma(\rho), & \eta < \rho.
    \end{array}
    \right.
    \label{ht4}
\end{equation}
By comparing the asymptotic formulas (\ref{ht2}) and (\ref{ht4}) we make sure
that at $\eta>\rho$ and $t\to \infty$ two leading terms of $h(t)$ can indeed be
determined from the modified Tauberian theorem. In contrast, if $\eta \leq
\rho$ then for finding two leading terms of the long-time expansion of $h(t)$
it is necessary to go beyond the Tauberian theorem.

\subsubsection{Properties of the diffusion coefficient.}
\label{sec:Coeff}

We define the coefficient of biased diffusion in the same way as for ordinary
diffusion:
\begin{equation}
    D_{\rm{b}}= \lim_{t\to \infty}\frac{\sigma^{2}_{X}(t)}{2t}.
    \label{Dud}
\end{equation}
As it follows from the modified Tauberian theorem and the definition (\ref{def
var}), for determining the law of biased diffusion, i.e., the long-time
behavior of the variance $\sigma^{2}_{X}(t)$, we should find two leading terms
of each Laplace transform, $\langle X_{t} \rangle_{s}$ and $\langle X_{t}^{2}
\rangle_{s}$, in the limit $s\to 0$. Assuming that the first two moments of the
residence time, $\overline{\tau}$ and $\overline{\tau^{2}}$, exist, the Laplace
transforms $p_{s}$ and $F_{s}$ at $s\to 0$ can be represented by the following
asymptotic formulas: $p_{s} \sim 1- s\overline{\tau} + s^{2}\overline{
\tau^{2}}/2$ and $F_{s} \sim 1/s - \overline{\tau}/2$. Substituting them into
(\ref{1sb}), we obtain the desired result
\begin{equation}
    \langle X_{t} \rangle_{s} \sim \frac{l}{\overline{\tau}}
    \frac{1}{s^{2}} + l \frac{\overline{\tau^{2}} -
    \overline{\tau}^{2}} {2\overline{\tau}^{2}} \frac{1}{s}.
    \label{<X1>s}
\end{equation}
Similarly, taking also into account that $G_{s} \sim 1/s$ as $s \to 0$, from
(\ref{2sb}) we find with the required accuracy
\begin{equation}
    \langle X_{t}^{2} \rangle_{s} \sim \frac{2l^{2}}
    {\overline{\tau}^{2}} \frac{1}{s^{3}} + 2l^{2}
    \frac{\overline{\tau^{2}} - \overline{\tau}^{2}}
    {\overline{\tau}^{3}} \frac{1}{s^{2}}.
    \label{<X2>s}
\end{equation}
The asymptotic formulas (\ref{<X1>s}) and (\ref{<X2>s}) are particular cases of
(\ref{hs2}) in which $L(1/s)$ is a constant. Therefore, the modified Tauberian
theorem leads to the following expressions for the moments $\langle X_{t}
\rangle$ and $\langle X_{t}^{2} \rangle$ in the long-time limit:
\begin{equation}
    \langle X_{t} \rangle \sim \frac{l}{\overline{\tau}}\,t
    + l \frac{\overline{\tau^{2}} - \overline{\tau}^{2}}
    {2\overline{\tau}^{2}},
    \quad
    \langle X_{t}^{2} \rangle \sim \frac{l^{2}} {\overline
    {\tau}^{2}}\,t^{2} + 2l^{2}\frac{\overline{\tau^{2}} -
    \overline{\tau}^{2}} {\overline{\tau}^{3}}\, t,
    \label{<X1,X2>}
\end{equation}
which in turn determine the law of biased diffusion
\begin{equation}
    \sigma^{2}_{X}(t) \sim l^{2}\frac{\overline{\tau^{2}} -
    \overline{\tau}^{2}} {\overline{\tau}^{3}}\,t.
    \label{var}
\end{equation}

Thus, it follows from (\ref{Dud}) and (\ref{var}) that the coefficient of
biased diffusion can be written as
\begin{equation}
    D_{\rm{b}}=  l^{2}\frac{\overline{\tau^{2}} -
    \overline{\tau}^{2}}{2\overline{\tau}^{3}}.
    \label{Dud1}
\end{equation}
Using (\ref{mth1}) and the Jensen inequality \cite{Fel}, one can see that for
all probability densities $u(g)$ the condition $D_{\rm{b}} \in [0,\infty]$
holds. However, if $f>g_{0}$ then the moments $\overline{\tau}$ and $\overline{
\tau^{2}}$ are finite (see section \ref{sec:Cond}) and nonzero (see formula
(\ref{mth1})), i.e., $D_{\rm{b}}\neq \infty$. On the other hand, representing
$\overline{ \tau^{2}}$ as
\begin{eqnarray}
    \overline{\tau^{2}} &=& \frac{(\nu l)^{2}}{2}\int_{-g_{0}}
    ^{g_{0}} \int_{-g_{0}}^{g_{0}} dgdg' u(g)u(g')
    \nonumber\\[6pt]
    &&\times \left( \frac{1}{(f+g)^{2}} + \frac{1}{(f+g')^{2}}\right)
    \label{rel3}
\end{eqnarray}
and $\overline{\tau}^{2}$ as
\begin{equation}
    \overline{\tau}^{2} = (\nu l)^{2}\int_{-g_{0}}^{g_{0}}
    \int_{-g_{0}}^{g_{0}}dgdg' \frac{u(g)u(g')}{(f+g)(f+g')},
    \label{rel4}
\end{equation}
we obtain
\begin{eqnarray}
    \overline{\tau^{2}} - \overline{\tau}^{2} &=& \frac{(\nu l)^{2}}{2}
    \int_{-g_{0}}^{g_{0}} \int_{-g_{0}}^{g_{0}} dgdg' u(g)u(g')
    \nonumber\\[6pt]
    &&\times \frac{(g-g')^{2}}{(f+g)^{2}(f+g')^{2}}.
    \label{rel5}
\end{eqnarray}
The last expression shows that always $\overline{\tau^{2}} -\overline{\tau}^{2}
\geq 0$, and the condition $\overline{\tau^{2}} = \overline{\tau}^{2}$
($D_{\rm{b}} =0$) holds only if $u(g) = \delta(g)$ (we took into account the
condition $u(-g) = u(g)$). But the probability density $u(g) = \delta(g)$
corresponds to the trivial case when the random force $g(x)$ is absent.
Excluding it from consideration, we obtain $D_{\rm{b}} \neq 0$. Thus, if
$f>g_{0}$ then $D_{\rm{b}} \in (0,\infty)$, i.e., the biased diffusion is
normal: $\sigma^{2}_{X}(t) \sim 2D_{\rm{b}}t$. It is clear from the above
analysis that at $f=g_{0}$ and $\alpha>2$ the biased diffusion is also normal.
In contrast, if $D_{\rm{b}}=0$ or $D_{\rm{b}} =\infty$ (these conditions are
realized only if $f=g_{0}$ and $\alpha \in (0,2]$, see section \ref{sec:Anom})
then the biased diffusion becomes anomalous: It is slower than normal in the
former case and is faster in the latter one. We note, however, that the
conditions $D_{\rm{b}}=0$ and $D_{\rm{b}} =\infty$ only indicate the existence
of anomalous behavior and cannot be used for finding the laws of anomalous
diffusion.

Next we consider the dependence of $D_{\rm{b}}$ on the external force $f$ and
statistical characteristics of the random force $g(x)$. Let us first determine
$D_{\rm{b}}$ in the case of large external force when $f\gg g_{0}$. By
expanding the integrand of (\ref{mth1}) in powers of $g/f$ and integrating over
$g$ with the condition $u(-g) = u(g)$, we obtain
\begin{equation}
    \overline{\tau} \sim \frac{\nu l}{f}\! \left( 1 + \frac{\sigma^{2}_{g}}
    {f^{2}} \right)\!,
    \quad
    \overline{\tau^{2}} \sim \left(\frac{\nu l}{f}\right)^{2}\!
    \left( 1 + 3\frac{\sigma^{2}_{g}} {f^{2}} \right)
    \label{moms1}
\end{equation}
as $f\to \infty$. The substitution of these asymptotic expressions into
(\ref{Dud1}) leads to the universal, inversely proportional dependence of
$D_{\rm{b}}$ on $f$ for large values of $f$:
\begin{equation}
    D_{\rm{b}} \sim \frac{l\sigma^{2}_{g}}{2\nu}\, \frac{1}{f}.
    \label{Dud2}
\end{equation}

\begin{figure}
    \centering
    \includegraphics[totalheight=4cm]{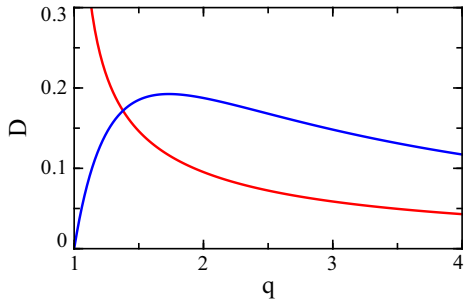}
    \caption{\label{fig.2} The reduced coefficient of unidirectional
    diffusion $D= D_{\rm{b}}\nu/lg_{0}$ as a function of the
    dimensionless driving force $q=f/g_{0}$. The monotonic
    dependence of $D$ on $q$ is determined from formula
    (\ref{Dud3}) which corresponds to the uniform probability
    density $u(g)$, and the non-monotonic one follows from
    formula (\ref{Dud5}) which corresponds to the probability
    density (\ref{u(g)2}).}
\end{figure}

If $f>g_{0}$ and the probability density $u(g)$ is given by (\ref{u(g)}) then,
according to (\ref{mth1}), the moments of the residence time can be represented
in the form
\begin{equation}
    \overline{\tau^{m}} = \left( \frac{\nu l}{f-g_{0}} \right)^{m}
    \!F\! \left( m,\beta;2\beta;-\frac{2g_{0}}{f-g_{0}}
    \right),
    \label{mth2}
\end{equation}
where $F(m,\beta;2\beta;z)=\!\,_{2}F_{1}(m,\beta;2\beta;z)$ is the
hypergeometric function. In some cases the first two moments, $\overline{\tau}$
and $\overline{\tau^{2}}$, can be expressed in elementary functions. In
particular, it is possible for $\beta=1$, i.e., if $u(g)$ corresponds to a
uniform distribution of the random force $g(x)$ on the interval $[-g_{0},
g_{0}]$. In this case, using the conditions \cite{AS} $F(1,1;2;z)= -z^{-1}\ln
(1-z)$, $F(m,\beta;2\beta;z)=F(\beta,m;2\beta;z)$ and $F(1,2;2;z)= (1-z)^{-1}$,
we get
\begin{equation}
    \overline{\tau} = \frac{\nu l}{2g_{0}}\, \ln \frac{q+ 1}{q- 1},
    \quad
    \overline{\tau^{2}} = \left(\frac{\nu l}{g_{0}}\right)^{2}
    \frac{1}{q^{2}-1}
    \label{moms2}
\end{equation}
and, as a consequence,
\begin{equation}
    D_{\rm{b}} = \frac{lg_{0}}{\nu}\! \left( \frac{4}{q^{2}-1} -
    \ln^{2} \frac{q+1}{q-1} \right) \ln^{-3} \frac{q+1}{q-1},
    \label{Dud3}
\end{equation}
where $q=f/g_{0}>1$ is the dimensionless external force. The coefficient of
biased diffusion (\ref{Dud3}) is a monotonically decreasing function of $q$
(see figure~\ref{fig.2}) which tends to infinity as $q\to 1$, $D_{\rm{b}} \sim
(2lg_{0}/\nu)(q-1)^{-1} |\ln^{-3} (q-1)|$, and approaches zero as $q\to
\infty$, $D_{\rm{b}} \sim (lg_{0}/6\nu) q^{-1}$. Since for the uniform
distribution $\sigma^{2}_{g}=g_{0}^{2}/3$, the last asymptotic formula
coincides with that given in (\ref{Dud2}).

Formula (\ref{mth2}) is also valid in the limit $f\to g_{0}$. However, in order
to find explicit expressions for $\overline{\tau}$ and $\overline {\tau^{2}}$,
it is much more convenient to use the basic formula (\ref{mth1}) with
$f=g_{0}$. A straightforward integration in this case yields
\begin{equation}
    \overline{\tau} = \frac{\nu l}{g_{0}}\,\frac{2\beta -1}
    {2(\beta - 1)},
    \quad
    \overline{\tau^{2}} = \left(\frac{\nu l}{g_{0}}\right)^{2}
    \frac{2\beta -1}{2(\beta - 2)}
    \label{moms3}
\end{equation}
for $\beta>1$ and $\beta>2$, respectively. Therefore, if $f=g_{0}$ (i.e.,
$q=1$) and $\beta>2$ then
\begin{equation}
    D_{\rm{b}} = \frac{lg_{0}}{\nu}\,\frac{\beta(\beta-1)}
    {(2\beta-1)^{2}(\beta-2)}.
    \label{Dud4}
\end{equation}

A quite different behavior of $D_{\rm{b}}$ on $q$ occurs if the probability
density $u(g)$ is mainly concentrated near the edges of the interval $[-g_{0},
g_{0}]$. In particular, for the limiting case characterized by the probability
density
\begin{equation}
    u(g) = \frac{1}{2}\, \delta(g-g_{0}) + \frac{1}{2}\, \delta(g+g_{0})
    \label{u(g)2}
\end{equation}
we obtain
\begin{equation}
    \overline{\tau} = \frac{\nu l}{g_{0}}\, \frac{q}{q^{2} - 1},
    \quad
    \overline{\tau^{2}} = \left(\frac{\nu l}{g_{0}}\right)^{2}
    \frac{q^{2}+1}{(q^{2}-1)^{2}}
    \label{moms4}
\end{equation}
and so
\begin{equation}
    D_{\rm{b}} = \frac{lg_{0}}{2\nu}\, \frac{q^{2}-1}{q^{3}}.
    \label{Dud5}
\end{equation}
In contrast to the previous case, now $D_{\rm{b}}$ is a non-monoto\-nic
function of $q$ (see figure \ref{fig.2}). According to (\ref{Dud5}),
$D_{\rm{b}} \sim (lg_{0}/2\nu)q^{-1}$ as $q\to \infty$ (since in this case
$\sigma^{2}_{g} = g^{2}_{0}$, the same result follows from (\ref{Dud2}) as
well), $D_{\rm{b}} \sim (lg_{0}/\nu)(q-1)$ as $q\to 1$, and the maximum value
of $D_{\rm{b}}$ is $D_{\rm{b}}\big|_{q=\sqrt{3}} \approx 0.192\! \times\!
(lg_{0}/\nu)$.

\subsection{Anomalous biased diffusion}
\label{sec:Anom}

According to the above analysis, the biased diffusion is anomalous if both
conditions $f=g_{0}$ and $\alpha \in (0,2]$ hold. Specifically, if $\alpha \in
(1,2]$ then $\overline{ \tau^{2}} = \infty$, $\overline{\tau}<\infty$ and so
$D_{\rm{b}}=\infty$, i.e., the biased diffusion is fast in comparison with
normal. In contrast, if $\alpha \in (0,1]$ then $\overline{ \tau^{2}} =
\infty$, $\overline{ \tau} = \infty$, and depending on $\alpha$ the biased
diffusion can be fast ($D_{\rm{b}}=\infty$), slow ($D_{\rm{b} }=0$) or even
formally normal ($0<D_{\rm{b}}<\infty$). Thus, taking into account that the
points $\alpha=1$ and $\alpha=2$ are distinct, we consider the long-time
behavior of the variance $\sigma^{2}_{X}(t)$ separately for $\alpha\in (0,1)$,
$\alpha\in (1,2)$, $\alpha = 1$, and $\alpha = 2$.

\subsubsection{$\alpha\in (0,1)$.}
\label{sec:Anom1}

Since in this case the leading terms of $\langle X_{t} \rangle^{2}$ and
$\langle X_{t}^{2} \rangle$ are different (see below), for their finding we
need to know only two leading terms of $p_{s}$ as $s\to 0$. These two terms can
be easily evaluated using the expression
\begin{equation}
    p_{s} = 1- \frac{1}{s}\int_{s\tau_{\rm{min}}}^{\infty} dy
    (1-e^{-y}) p\!\left( \frac{y}{s} \right),
    \label{def1}
\end{equation}
which follows from the definition $p_{s}= \int_{\tau_{\rm{min}}} ^{\infty}
d\tau e^{-s\tau}p(\tau)$ and the normalization condition $\int_{\tau_{
\rm{min}}} ^{\infty} d\tau p(\tau)=1$ (we recall that $\tau_{\rm{min}} = \nu
l/2g_{0}$ and $\tau_{\rm{max}} = \infty$ at $f=g_{0}$). Keeping in (\ref{def1})
the leading term of the asymptotic expansion of the integral as $s\to 0$ and
using  asymptotic formula (\ref{asymp}) and the integral representation of the
gamma function \cite{AS}, $\Gamma(x) = \int_{0}^{\infty}dy\, e^{-y} y^{x - 1}$,
we obtain the desired result
\begin{equation}
    p_{s} \sim 1- \frac{a\Gamma(1-\alpha)}{\alpha}\,s^{\alpha}.
    \label{ps1}
\end{equation}

In order to find the leading terms of $F_{s}$ and $G_{s}$, it is convenient to
introduce the new variable of integration $y=s\tau$ and rewrite expressions
(\ref{F}) and (\ref{G}) in the form
\begin{equation}
    F_{s} = \frac{1}{s} - \frac{1}{s^{2}}\int_{s\tau_{\rm{min}}}^
    {\infty}dy\,\frac{e^{-y} + y -1} {y}\,p\!\left(
    \frac{y}{s} \right)
    \label{F1}
\end{equation}
and
\begin{equation}
    G_{s} = \frac{1}{s} - \frac{2}{s^{2}}\int_{s\tau_{\rm{min}}}^
    {\infty}dy\,\frac{(1+y)e^{-y} + y^{2}/2 -1}{y^{2}}\,p\!\left(
    \frac{y}{s} \right).
    \label{G1}
\end{equation}
Using the asymptotic formula (\ref{asymp}), one can make sure that the integral
terms in (\ref{F1}) and (\ref{G1}) are proportional to $1/s^{1- \alpha}$ and so
$F_{s} \sim G_{s} \sim 1/s$ as $s\to 0$. Thus, in accordance with this result
and asymptotic expression (\ref{ps1}) the leading terms of the Laplace
transforms (\ref{1sb}) and (\ref{2sb}) take the form
\begin{equation}
    \langle X_{t} \rangle_{s} \sim \frac{l\alpha}{a\Gamma(1-\alpha)}
    \frac{1}{s^{1+\alpha}},
    \quad
    \langle X_{t}^{2} \rangle_{s} \sim \frac{2l^{2}\alpha^{2}}
    {a^{2}\Gamma^{2}(1-\alpha)}\frac{1}{s^{1+2\alpha}}.
    \label{<X1,X2>s1}
\end{equation}

These asymptotic formulas are particular cases of the asymptotic formula
(\ref{hs1}) in which the slowly varying function $L(1/s)$ is a constant.
Therefore, as follows from (\ref{ht1}), the long-time dependence of the moments
$\langle X_{t} \rangle$ and $\langle X_{t}^{2} \rangle$ is given by
\begin{equation}
    \langle X_{t} \rangle \sim \frac{l\alpha}{a\Gamma(1-\alpha)
    \Gamma(1+\alpha)}\,t^{\alpha}
    \label{<X1>1}
\end{equation}
and
\begin{equation}
    \langle X_{t}^{2} \rangle \sim \frac{2l^{2}\alpha^{2}}
    {a^{2}\Gamma^{2}(1-\alpha)\Gamma(1+2\alpha)}\,t^{2\alpha}.
    \label{<X2>1}
\end{equation}
As it seen from these asymptotic formulas, the leading terms of $\langle X_{t}
\rangle^{2}$ and $\langle X_{t}^{2} \rangle$ are different and so
\begin{equation}
    \sigma^{2}_{X}(t) \sim \frac{l^{2}\alpha^{2}}{a^{2}\Gamma^{2}
    (1-\alpha)} \left(\! \frac{2} {\Gamma(1+2\alpha)} - \frac{1}
    {\Gamma^{2}(1+\alpha)}\right)\! t^{2\alpha}.
    \label{var1}
\end{equation}
According to (\ref{var1}), the biased diffusion is characterized by a
subdiffusive behavior (with $D_{\rm{b}}=0$) if $\alpha\in (0,1/2)$ and by a
superdiffusive one (with $D_{\rm{b}}=\infty$) if $\alpha\in (1/2,1)$. At
$\alpha=1/2$ the diffusion is formally normal with $D_{\rm{b}}= l^{2}(\pi -
2)/(2\pi a)^{2}$. However, since $\overline{ \tau} = \infty$, $\overline{
\tau^{2}} = \infty$ and $\langle X_{t} \rangle \propto t^{1/2}$, we term this
type of biased diffusion the quasi-normal diffusion. We note also that the
asymptotic formula (\ref{var1}) agrees with the asymptotic solution of the
CTRWs \cite{Shles}.

\subsubsection{$\alpha\in (1,2)$.}
\label{sec:Anom2}

In this and all other cases the leading terms of $\langle X_{t} \rangle^{2}$
and $\langle X_{t}^{2} \rangle$ are the same. Therefore, for finding
$\sigma^{2}_{X}(t)$ we need to determine three leading terms of $p_{s}$ as $s
\to 0$. Since for these values of $\alpha$ the mean residence time $\overline
{\tau} = \int_{\tau_{\rm{min}}}^ {\infty} d\tau \tau p(\tau)$ exists, to this
end it is convenient to use the following exact formula:
\begin{equation}
    p_{s} = 1- \overline{\tau}s + \frac{1}{s} \int_{s\tau_{\rm{min}}}
    ^{\infty} dy (e^{-y} + y -1) p\!\left( \frac{y}{s} \right).
    \label{def2}
\end{equation}
Proceeding in the same way as before, at $s\to 0$ we obtain
\begin{equation}
    p_{s} \sim 1 - \overline{\tau}s + \frac{a\Gamma(2-\alpha)}
    {\alpha(\alpha - 1)}\,s^{\alpha}.
    \label{ps2}
\end{equation}
Then, from (\ref{F1}) and (\ref{G1}) it follows that $F_{s} \sim 1/s -
\overline{\tau}/2$ and $G_{s} \sim 1/s - 2\overline{\tau}/3$ as $s\to 0$.
Taking also into account that in the main approximation $F_{s} \sim 1/s$ and
$2p_{s} F_{s} + (1-p_{s})G_{s} \sim 2/s$, the Laplace transforms (\ref{1sb})
and (\ref{2sb}) are reduced to
\begin{equation}
    \langle X_{t} \rangle_{s} \sim \frac{l}{\overline{\tau}}
    \frac{1}{s^{2}} + \frac{la\Gamma(2-\alpha)}{\overline{\tau}^{2}
    \alpha(\alpha -1)} \frac{1}{s^{3-\alpha}}
    \label{<X>s2}
\end{equation}
and
\begin{equation}
    \langle X_{t}^{2} \rangle_{s} \sim \frac{2l^{2}}{\overline
    {\tau}^{2}} \frac{1}{s^{3}} + \frac{4l^{2}a\Gamma(2-\alpha)}
    {\overline{\tau}^{3}\alpha(\alpha -1)}\frac{1}{s^{4-\alpha}}.
    \label{<X2>s2}
\end{equation}

Finally, applying the modified Tauberian theorem to (\ref{<X>s2}) and
(\ref{<X2>s2}) and using the well-known property of the gamma function,
$\Gamma(1+x) = x\Gamma(x)$, we get
\begin{equation}
    \langle X_{t} \rangle \sim \frac{l}{\overline{\tau}}\,t +
    \frac{la}{\overline{\tau}^{2}\alpha(\alpha -1)(2-\alpha)}\,
    t^{2-\alpha}
    \label{<X>2}
\end{equation}
and
\begin{equation}
    \langle X_{t}^{2} \rangle \sim \frac{l^{2}}{\overline{\tau}^{2}}
    \,t^{2} + \frac{4l^{2}a}{\overline{\tau}^{3}\alpha(\alpha -1)
    (2-\alpha)(3-\alpha)}\, t^{3-\alpha}.
    \label{<X2>2}
\end{equation}
as $t\to \infty$. As it follows from (\ref{<X>2}) and (\ref{<X2>2}), the
long-time behavior of the variance $\sigma^{2}_{X}(t)$ obeys the asymptotic
power law
\begin{equation}
    \sigma^{2}_{X}(t) \sim \frac{2l^{2}a}{\overline{\tau}^{3}\alpha
    (2-\alpha)(3-\alpha)}\, t^{3-\alpha}.
    \label{var2}
\end{equation}
It shows that at $\alpha \in (1,2)$ the unidirectional transport of particles
is superdiffusive. The parameters in (\ref{var2}) depend on the probability
density $u(g)$ of the random force. In particular, if $u(g)$ is determined by
(\ref{u(g)}) then $\alpha = \beta$, the parameter $a$ is given by (\ref{a}),
and
\begin{equation}
    \overline{\tau} = \frac{\nu l}{g_{0}}\, \frac{\alpha -1/2}{\alpha -1}.
    \label{mean tau}
\end{equation}

\subsubsection{$\alpha=1$.}
\label{sec:Anom3}

In order to illustrate the distinctive features of the long-time dependence of
the moments $\langle X_{t} \rangle$ and $\langle X_{t}^{2} \rangle$ at $\alpha=
1$, i.e., when according to (\ref{asymp}) $p(\tau) \sim a/\tau^{2}$ ($\tau \to
\infty$), we first represent the Laplace transform $p_{s}$ in the form
\begin{equation}
    p_{s} = 1- aq_{s} - \frac{1}{s} \int_{s\tau_{\rm{min}}}^
    {\infty} dy (1-e^{-y})\! \left[ p\!\left( \frac{y}{s} \right)
    - a\frac{s^{2}}{y^{2}} \right]\!,
    \label{def3}
\end{equation}
where $q_{s} = s\int_{s\tau_{\rm{min}}}^{\infty} dy (1-e^{-y})/ y^{2}$. With
the definition of the exponential integral \cite{AS}, $E_{1}(x) =\int_{x}^
{\infty} dy\,e^{-y}/y$, the last quantity can be written as
\begin{equation}
    q_{s} = \frac{1-e^{-s\tau_{\rm{min}}}}{\tau_{\rm{min}}} +
    sE_{1}(s\tau_{\rm{min}}).
    \label{qs1}
\end{equation}
Then, since $p(\tau) - a/\tau^{2} = o(1/\tau^{2})$ as $\tau \to \infty$ (we
recall that $p(\tau) \sim a/\tau^{2}$ if $f=g_{0}$ and $\alpha = 1$), the
integral term in (\ref{def3}) at $s\to 0$ can be neglected in comparison with
the term $aq_{s}$. Finally, using the asymptotic formula $E_{1}(s\tau_
{\rm{min}}) \sim \ln (1/s)$ ($s\to 0$) \cite{AS}, we obtain
\begin{equation}
    p_{s} \sim 1 - as\ln \frac{1}{s}.
    \label{ps3}
\end{equation}

As it can be easily seen, the integral terms in (\ref{F1}) and (\ref{G1}) at
$\alpha = 1$ are proportional to $\ln (1/s)$ and so $F_{s} \sim G_{s} \sim 1/s$
($s\to 0$). Therefore, the leading terms of the Laplace transforms (\ref{1sb})
and (\ref{2sb}) can be written in the form
\begin{equation}
    \langle X_{t} \rangle_{s} \sim  \frac{l}{a}\frac{1}{s^{2}
    \ln(1/s)},
    \quad
    \langle X_{t}^{2} \rangle_{s} \sim \frac{l^2}{a^2}
    \frac{2}{s^{3}\ln^{2}(1/s)}
    \label{X(1,2)s3}
\end{equation}
($s\to 0$) that in accordance with (\ref{ht1}) yields
\begin{equation}
    \langle X_{t} \rangle \sim  \frac{l}{a}\frac{t}{\ln t},
    \quad
    \langle X_{t}^{2} \rangle \sim \frac{l^2}{a^2}
    \frac{t^2}{\ln^{2} t}
    \label{X(1,2)3}
\end{equation}
($t\to \infty$). Since $\langle X_{t} \rangle^{2} \sim \langle X_{t}^{2}
\rangle$, for finding the long-time behavior of the variance $\sigma^{2}_{X}
(t)$ we need to know at least two terms of the asymptotic expansion of $\langle
X_{t} \rangle$ and $\langle X_{t}^{2} \rangle$. However, in contrast to the
previous case they cannot be determined from the modified Tauberian theorem
because the leading terms of $\langle X_{t} \rangle_{s}$ and $\langle X_{t}^{2}
\rangle_{s}$ contain the slowly varying functions, see (\ref{X(1,2)s3}). As it
was mentioned in section \ref{sec:Taub}, in this case it is necessary to go
beyond the Tauberian theorem. Nevertheless, using the asymptotic laws
(\ref{var1}) and (\ref{var2}), we can make some qualitative conclusions about
the character of biased diffusion at $\alpha = 1$. First of all, it follows
from (\ref{var1}) and (\ref{var2}) that $t^{2\alpha}$ and $t^{3-\alpha}$ tend
to $t^{2}$ as $\alpha \to 1-0$ and $\alpha \to 1+0$, respectively. But since
$\Gamma(1-\alpha) \to \infty$ as $\alpha \to 1-0$ and $\overline{\tau} \to
\infty$ as $\alpha \to 1+0$, the coefficients of proportionality between
$\sigma^{2}_{X}(t)$ and $t^{2}$ tend to zero. This means that at $\alpha = 1$
the weakened ballistic diffusion should occur.

\subsubsection{$\alpha=2$.}
\label{sec:Anom4}

If $\alpha=2$ then it is convenient to represent the Laplace transform $p_{s}$
as
\begin{equation}
    p_{s} = 1- \overline{\tau}s - ar_{s} - \frac{1}{s}\!
    \int_{s\tau_{\rm{min}}}^ {\infty}\! dy (1- y -e^{-y})\!
    \left[ p\!\left( \frac{y}{s}\right) -
    a\frac{s^{3}}{y^{3}} \right]\!,
    \label{def4}
\end{equation}
where
\begin{eqnarray}
    r_{s} &=& s^{2}\int_{s\tau_{\rm{min}}}^{\infty} dy\,
    \frac{1- y - e^{-y}}{y^{3}}
    \nonumber\\[6pt]
    &=& \frac{1 - 2s\tau_{\rm{min}}\! -
    (1 - s\tau_{\rm{min}}) e^{-s\tau_{\rm{min}}}}
    {2\tau_{\rm{min}}^{2}} - \frac{s^{2}}
    {2}E_{1}(s\tau_{\rm{min}}).\qquad
    \label{ws}
\end{eqnarray}
Taking into account that $p(\tau) - a/\tau^{3} = o(1/\tau^{3})$ ($\tau \to
\infty$), the integral term in (\ref{def4}) at $s\to 0$ can be neglected in
comparison with $ar_{s} \sim -(as^{2}/2) \ln(1/s)$. Therefore, keeping in
$p_{s}$ three leading terms, one obtains
\begin{equation}
    p_{s} \sim 1 - \overline{\tau}s + \frac{a}{2}s^{2}\ln \frac{1}{s}.
    \label{ps4}
\end{equation}

Asymptotic expression (\ref{ps4}) shows that for determining two leading terms
of $\langle X_{t} \rangle_{s}$ and $\langle X_{t}^{2} \rangle_{s}$ only the
leading terms of $F_{s}$ and $G_{s}$ should be kept. Since $F_{s} \sim G_{s}
\sim 1/s$, from (\ref{1sb}), (\ref{2sb}) and (\ref{ps4}) we immediately find
\begin{equation}
    \langle X_{t} \rangle_{s} \sim \frac{l}{\overline{\tau}}
    \frac{1}{s^{2}} + \frac{la} {2\overline{\tau}^{2}}
    \frac{1}{s}\ln\! \frac{1}{s},
    \quad
    \displaystyle \langle X_{t}^{2} \rangle_{s} \sim \frac{2l^{2}}
    {\overline{\tau}^{2}} \frac{1}{s^{3}} + \frac{2l^{2}a}
    {\overline{\tau}^{3}}\frac{1}{s^{2}} \ln\! \frac{1}{s}
    \label{X(1,2)s4}
\end{equation}
($s\to 0$). Accordingly, the modified Tauberian theorem, see (\ref{hs2}) and
(\ref{ht2}), leads to
\begin{equation}
    \displaystyle \langle X_{t} \rangle \sim \frac{l}{\overline
    {\tau}}\, t + \frac{la}{2\overline{\tau}^{2}}\ln t,
    \quad
    \displaystyle \langle X_{t}^{2} \rangle \sim \frac{l^{2}}
    {\overline{\tau}^{2}} \,t^{2} + \frac{2l^{2}a}
    {\overline{\tau}^{3}}\, t\ln t
    \label{X(1,2)4}
\end{equation}
($t\to \infty$) and so the variance of the particle position in the long-time
limit is described by the asymptotic formula
\begin{equation}
    \sigma^{2}_{X}(t) \sim \frac{l^{2}a} {\overline{\tau}^{3}}\, t\ln t.
    \label{var4}
\end{equation}
It shows that the biased diffusion at $\alpha =2$ is logarithmically enhanced
in comparison with normal diffusion occurring at $\alpha>2$. The fact that
$\sigma^{2}_{X}(t)$ at $t\to \infty$ increases faster than $t$ is in accordance
with the diffusion law (\ref{var2}). Indeed, while $t^{3-\alpha} \to t$ as
$\alpha \to 2-0$, the coefficient of proportionality between the variance and
time tends to infinity.

\begin{figure}
    \centering
    \includegraphics[totalheight=4cm]{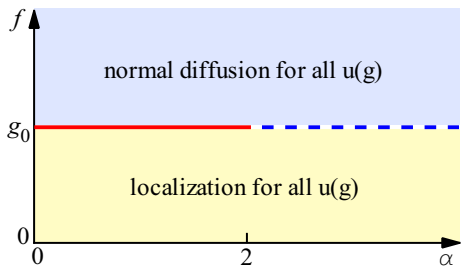}
    \caption{\label{fig.3} Regimes of biased diffusion in a
    piecewise linear random potential. At $f>g_{0}$ the normal
    biased diffusion occurs for all symmetric probability
    densities $u(g)$ with support on the interval $[-g_{0},g_{0}]$.
    A limiting case of subdiffusion, i.e., localization of particles
    at the average distance (\ref{aver}) from the origin, is
    realized at $f<g_{0}$. If $f=g_{0}$ then the character of
    diffusion depends on the exponent $\alpha$ characterizing the
    asymptotic behavior (\ref{asymp}) of the residence time
    probability density $p(\tau)$. Specifically, normal diffusion
    occurs at $\alpha>2$ (dashed line) and anomalous one at
    $\alpha\in (0,2]$ (solid line).
    In turn, subdiffusion occurs if $\alpha \in (0,1/2)$,
    superdiffusion if $\alpha \in (1/2,1)$ or $\alpha \in (1,2)$,
    and other types of anomalous diffusion are realized at
    $\alpha=1/2$ (quasi-normal diffusion), $\alpha =1$ (weakened
    ballistic diffusion), and $\alpha =2$ (logarithmically
    enhanced normal diffusion).}
\end{figure}

For convenience, the above considered regimes of biased diffusion that occurs
under a constant force in a piecewise linear random potential are summarized in
figure \ref{fig.3}.
We note in this context that all laws of diffusion were obtained within a
single model dealing with the solution $X_{t}$ of the motion equation (\ref{eq
motion}). In contrast, the CTRW jump model which deals with the discrete
variable $Y_{t}$ reproduces only the long-time behavior of the variance
$\sigma^{2}_{X}(t)$ if the waiting time is associated with the residence time
\cite{DK}. At short times the behavior of $\sigma^{2}_{X}(t)$ and $\sigma^{2}_
{Y}(t)$ is completely different, see sections \ref{sec:Short}.

We complete our analysis of biased diffusion by comparing the root-mean-square
displacement $\sigma_{X}(t)$ and the average displacement $\langle X_{t}
\rangle$ at long times. Since $\sigma_{X}(t)$ characterizes the spreading of
particles around their mean position $\langle X_{t} \rangle$, the coefficient
of variation $C(t)= \sigma_{X}(t)/\langle X_{t} \rangle$ can be considered as a
relative measure of the intensity of biased diffusion. Using the previous
results, for normal diffusion we obtain $C(t) \propto 1/t^{1/2}$ ($t\to
\infty$). In the case of anomalous diffusion the coefficient of variation at
$\alpha \in [1,2]$ also tends to zero in the long-time limit, though more
slowly than $1/t^{1/2}$. In particular, if $\alpha \in (1,2)$ then, according
to (\ref{<X>2}) and (\ref{var2}), $C(t) \propto 1/t^{(\alpha -1 )/2}$. In
contrast, as follows from (\ref{<X1>1}) and (\ref{var1}), the coefficient of
variation $C(\infty)$ at $\alpha \in (0,1)$ does not vanish:
\begin{equation}
    C(\infty) = \! \left( \frac{2\Gamma^{2}(1+\alpha)}
    {\Gamma(1+2\alpha)} - 1 \right)^{\!1/2}.
    \label{C}
\end{equation}
In this case the biased diffusion is so intensive that the spreading of
particles is of the order of their displacement, i.e., $\sigma_{X} (t) \sim
C(\infty) \langle X_{t} \rangle$. Interestingly, since $C(\infty) \to 0$ as
$\alpha \to 1$ and $C(\infty) \to 1$ as $\alpha \to 0$, the slower the
diffusion, the larger its intensity.

\section{Conclusions}
\label{sec:Concl}

We have studied the unidirectional transport of particles which occurs under a
constant external force in a piecewise linear random potential. The slopes of
this potential, i.e., realizations of the piecewise constant random force, are
assumed to be independent on different intervals of a fixed length and
distributed with the same probability density. Using the overdamped motion
equation, we have derived the probability density of the particle position and
have calculated its first two moments by the Laplace transform method. The
Laplace transforms of these moments are represented by means of the probability
density of the residence time, i.e., time that a particle spends moving on the
interval of a fixed length, which in turn is explicitly expressed through the
probability density of the random force.

It has been shown that if the external force is less than the upper bound of
the random force then the limiting case of subdiffusion, i.e., particle
localization, occurs. In this regime, we have calculated the average distance
between the origin and the points of localization which is always finite. In
contrast, if the external force exceeds the upper bound of the random force
then particles can be transported to an arbitrary large distance. Because of
the influence of the random force, in this case the unidirectional motion of
particles has a diffusive character. We have shown by applying the ordinary
Tauberian theorem for the Laplace transform that at small times the biased
diffusion is always ballistic. For the analysis of the transport properties at
long times we used the modified Tauberian theorem that in most cases permits us
to find the two leading terms of the asymptotic expansion of the first and
second moments of the particle position. Within this approach, we have shown
that the biased diffusion is normal and the diffusion coefficient as a function
of the external force can be either monotonic or non-monotonic. The latter
occurs if the probability density of the random force is concentrated near the
edges of the interval of support.

If the external force is equal to the boundary value of the random force then
at short times the biased diffusion remains ballistic, but at long times it can
be either normal or anomalous. In the last case the character of diffusion
depends on the asymptotic behavior of the probability density of the residence
time, which is described by the exponent $\alpha$. Using the modified Tauberian
theorem, it has been shown that at $\alpha \in (0,2]$ the biased diffusion is
anomalous. Specifically, subdiffusion, i.e., power-law dependence of the
variance on time with power less than 1 occurs at $\alpha \in (0,1/2)$, and
superdiffusion, i.e., power-law dependence of the variance on time with power
greater than $1$ occurs at $\alpha \in (1/2,1)$ and $\alpha \in (1,2)$. The
quasi-normal diffusion, which is characterized by the nonlinear time dependence
of the first moment of the particle position and the linear dependence of the
variance, is realized at $\alpha = 1/2$. Finally, the weakened ballistic
diffusion and the logarithmically enhanced normal diffusion are realized at
$\alpha = 1$ and $\alpha = 2$, respectively.


\begin{thebibliography}{99}

\bibitem{BG} J.P. Bouchaud, A. Georges, Phys. Rep. {\bf 195}, 127 (1990)

\bibitem{Sch} S. Scheidl, Z. Phys. B {\bf 97}, 345 (1995)
\bibitem{DV} P. Le Doussal, V.M. Vinokur, Physica C {\bf 254}, 63 (1995)
\bibitem{PKDK} P.E. Parris, M. Ku\'{s}, D.H. Dunlap, V.M. Kenkre, Phys.
    Rev. E {\bf 56}, 5295 (1997)
\bibitem{GB} D.A. Gorokhov, G. Blatter, Phys. Rev. B {\bf 58}, 213 (1998)
\bibitem{DH} S.I. Denisov, W. Horsthemke, Phys. Rev. E {\bf 62}, 3311 (2000)
\bibitem{LV} A.V. Lopatin, V.M. Vinokur, Phys. Rev. Lett. {\bf 86}, 1817 (2001)
\bibitem{Mon} C. Monthus, Lett. Math. Phys. {\bf 78}, 207 (2006)
\bibitem{RE} P. Reimann, R. Eichhorn, Phys. Rev. Lett. {\bf 101}, 180601 (2008)

\bibitem{PASF} M.N. Popescu, C.M. Arizmendi, A.L. Salas-Brito, F. Family,
    Phys. Rev. Lett. {\bf 85}, 3321 (2000)
\bibitem{GLZH} L. Gao, X. Luo, S. Zhu, B. Hu, Phys. Rev. E {\bf 67}, 062104
    (2003)
\bibitem{ZLAF} D.G. Zarlenga, H.A. Larrondo, C.M. Arizmendi, F. Family,
    Phys. Rev. E {\bf 75}, 051101 (2007)
\bibitem{DLDHK} S.I. Denisov, T.V. Lyutyy, E.S. Denisova, P. H\"{a}nggi, H.
    Kantz, Phys. Rev. E {\bf 79}, 051102 (2009)

\bibitem{KLS} H. Kunz, R. Livi, A. S\"{u}t\H{o}, Phys. Rev. E {\bf 67},
    011102 (2003)
\bibitem{DKDH1} S.I. Denisov, M. Kostur, E.S. Denisova, P. H\"{a}nggi,
    Phys. Rev. E {\bf 75}, 061123 (2007)
\bibitem{DKDH2} S.I. Denisov, M. Kostur, E.S. Denisova, P. H\"{a}nggi,
    Phys. Rev. E {\bf 76}, 031101 (2007)

\bibitem{DK} S.I. Denisov, H. Kantz, Phys. Rev. E {\bf 81}, 021117 (2010)

\bibitem{MW} E.W. Montroll, G.H. Weiss, J. Math. Phys. {\bf 6}, 167 (1965)

\bibitem{Hug} B.D. Hughes, {\it Random Walks and Random Environments }
    (Clarendon Press, Oxford, 1995), Vol. 2
\bibitem{AH} D. ben-Avraham, S. Havlin, {\it Diffusion and Reactions
    in Fractals and Disordered Systems} (Cambridge University Press, Cambridge,
    2000)
\bibitem{MK} R. Metzler, J. Klafter, Phys. Rep. {\bf 339}, 1 (2000)
\bibitem{Z} G. Zaslavsky, Phys. Rep. {\bf 371}, 461 (2002)
\bibitem{KRS} {\it Anomalous Transport: Foundations and Applications}, edited
    by R. Klages, G. Radons, I.M. Sokolov (Wiley-VCH, Berlin, 2008)

\bibitem{SWK} M.F. Shlesinger, B. West, J. Klafter, Phys. Rev. Lett. {\bf
    58}, 1100 (1987)
\bibitem{MLW} J. Masoliver, K. Lindenberg, G.H. Weiss, Physica A {\bf 157},
    891 (1989)
\bibitem{ZK} G. Zumofen, J. Klafter, Phys. Rev. E {\bf 47}, 851 (1993)
\bibitem{Bar} E. Barkai, Chem. Phys. {\bf 284}, 13 2002
\bibitem{SM} I.M. Sokolov, R. Metzler, Phys. Rev. E {\bf 67}, 010101(R) (2003)

\bibitem{ZSS} V. Zaburdaev, M. Schmiedeberg, H. Stark, Phys. Rev. E {\bf 78},
    011119 (2008)

\bibitem{Fel} W. Feller, {\it An Introduction to Probability Theory
    and its Applications} (Wiley, New York, 1971), Vol. 2

\bibitem{GK} B.V. Gnedenko, V.Yu. Korolev, {\it Random Summation:
    Limit Theorems and Applications} (CRC Press, Boca Raton, FL, 1996)

\bibitem{Erd} A. Erd\'{e}lyi, {\it Tables of Integral Transforms} (McGraw-Hill,
    New York, 1954), Vol. 1

\bibitem{AS} M. Abramowitz, I.A. Stegun, {\it Handbook of Mathematical
    Functions} (Dover, New York, 1972)

\bibitem{Shles} M.F. Shlesinger, J. Stat. Phys. {\bf 10}, 421 (1974)

\end{thebibliography}
\end{document}